\documentclass[12pt,twoside,a4paper,openright]{article}
\usepackage[dvips]{epsfig}
\usepackage[english]{babel}
\usepackage[]{caption}
\begin{document}
\title{LCFI Vertex Package\\
{\small Presented at the 17th International Workshop on Vertex detectors}\\[-1.0ex]
{\small July 28 - 1 August  2008, Ut\"{o} Island, Sweden}}

\author{Sonja Hillert\\ {\small on behalf of the LCFI Collaboration} 
\footnote{\copyright Copyright owned by the author(s) under the terms of the
Creative Commons Attribution-Noncommercial-ShareAlike License.}
\\[2ex]
     {\it\small University of Oxford - Department of Physics}\\
     {\it\small Denys Wilkinson Building, Keble Road, Oxford OX1 3RH - UK}\\
     {\it\small E-mail: s.hillert1@physics.ox.ac.uk}}

\date{}

\maketitle

\abstract{The LCFIVertex software, developed by the Linear Collider
Flavour Identification (LCFI) collaboration and providing tools for
vertexing, flavour tagging and quark charge determination for 
low-mass vertex detectors of high point resolution is presented. 
Particular emphasis is given to code extensions since the first 
release in April 2007.
A recently developed new vertex finder, ZVMST, and its performance 
at $\sqrt{s} = 91.2\,\mathrm{GeV}$, are described in more detail.}
%
%
\section{Introduction}
\label{Introduction}
The International Linear Collider (ILC), a $200$ - $500\,\mathrm{GeV}$ 
$e^{+}e^{-}$ collider, is envisaged by the particle physics community as next major 
accelerator following the LHC \cite{Behnke:2001qq,RDR:2007}. 
The high precision measurements planned at this machine place demanding constraints on the 
quality of the detectors as well as requiring excellent software for event reconstruction 
and data analysis. 
In the context of vertex detector R\&D for the ILC, and more generally applicable for 
low-mass vertex detectors of high point resolution, software for vertexing, flavour tagging 
and quark charge determination has been developed. 
This code named LCFIVertex \cite{LCFIVertexPaperInPrep,LCFIVertexDocumentation} and 
interfaced to one of the ILC software frameworks, is currently widely used by both the 
ILD \cite{ILD} and SiD \cite{SiD} detector concept groups for detector optimisation and 
preparation of Letters of Intent due in spring 2009.

This paper is organised as follows: section \ref{FrameworkAndDetector} gives an overview of 
the framework to which the code is interfaced and a typical detector model used in the 
examples of code performance. 
An overview of core functionality of the code and of updates since the first release is given 
in section \ref{LCFIVertexOverview}. 
One of the recent extensions of the package, the minimum spanning tree-based vertex finder 
ZVMST \cite{Hillert:2008}, is described in more detail in section \ref{ZVMST}. 
It is followed by a preliminary performance comparison between ZVMST and the leading vertex 
finder \verb|ZVTOP_ZVRES| in section \ref{PerformanceComparison}. 
Section \ref{SummaryAndConclusions} summarises the main results.
%
%
\section{Marlin framework and vertex detector used in performance studies}
\label{FrameworkAndDetector}
The LCFIVertex code is based on the event data model LCIO \cite{Gaede:2003ip} permitting 
the exchange of Monte Carlo (MC) samples between the different software frameworks in 
use by the ILC community, and the modular analysis framework Marlin \cite{Gaede:2006}, 
facilitating distributed code development. 

The MC program Pythia was used to generate events, which were passed through 
the GEANT4-based \cite{GEANT4:2003:2006} detector simulation MOKKA 
\cite{Mora de Freitas:2004sq} using the \verb|LDC01_05Sc| detector model 
\cite{LDC01_05Sc}. 
Events were reconstructed using the MarlinReco package \cite{Wendt:2007iw}, 
in particular the digitization and \verb|FullLDCTracking| code by A. Raspereza 
\cite{Raspereza:2008} and the particle flow package PandoraPFA by M. Thomson 
\cite{Thomson:2006}. 

In the detector model \verb|LDC01_05Sc|, developed by the LDC detector concept group, 
now part of the ILD group, the vertex detector design envisages 5 layers of sensors 
with radial positions, numbers of sensors in the layers and sensor dimensions as
described in the TESLA TDR \cite{Behnke:2001qq}, i.e. five evenly spaced layers
with the innermost radius being $15\,\mathrm{mm}$ and that of the outermost layer
being $60\,\mathrm{mm}$. 
Sensors and support together are assumed to correspond to a material budget of 
$0.1\% X_{0}$ per layer. 
The point resolution used in the MC test sample is $2.8\,\mu m$, as currently 
used in the benchmark studies performed by the ILD detector concept group. 
Since the forward region of the detector is of particular importance at the ILC, and 
hermeticity of all systems is hence crucial, vertex detector R\&D groups aim at a 
polar angle coverage of $|\mathrm{cos}\,\theta| \leq 0.96$.
The $B$-field is assumed to be $4\,\mathrm{T}$.
%
%
\section{Overview of the LCFIVertex software}
\label{LCFIVertexOverview}
The first version of the LCFIVertex code was released in April 2007 \cite{Hillert:2007}. 
In developing this new C++ based code the LCFI physics group has been building on earlier 
work by the SLD collaboration \cite{Jackson:1996sy,Thom:2002} 
and by LCFI and the TESLA detector R\&D group 
\cite{Hawkings:2000,Xella-Hansen:2001,Xella-Hansen:2003,Behnke:2001qq}, permitting 
detailed code validation using the fast MC program SGV 
\cite{SGV} for which an interface to a FORTRAN implementation of the core algorithms 
was available \cite{Adler:2006dw}.

The core modules of LCFIVertex, all run on a jet-by-jet basis, provide the vertex finder 
\verb|ZVTOP|, a neural-net based flavour tag following the method developed by R.~Hawkings and 
quark charge reconstruction. 
The \verb|ZVTOP| vertex finder comprises three algorithms, each based on a different ansatz 
for finding decay vertices of heavy flavour hadrons from the topological information contained 
in the tracks of the input jet. 
The leading algorithm most widely used in the ILC community is the \verb|ZVRES| approach; 
its name refers to the fact that the algorithm strongly relies on a criterion for checking 
whether different vertices are resolved from each other. 
In contrast, the \verb|ZVKIN| or ghost track algorithm is a more specialised approach dedicated 
to $b$-jets, which approximates the direction of flight of the decayed $B$-hadron and uses 
this additional kinematic information to identify the decay vertices. 
This approach permits vertices to be found also in jets, in which both the $B$- and the 
subsequent charm decay are one-pronged. 
The new ZVMST approach resembles \verb|ZVRES| in scope and mathematical description of the 
topological information but uses a minimum spanning tree (MST) to select the most promising 
vertex candidates. 

Observables based on secondary vertices, such as the $P_T$-corrected vertex mass 
\cite{SLD:RbUsingVertexMassTag:1998}, the secondary vertex probability and its decay length, 
provide the most stringent indication of the jet flavour. 
Such variables therefore form the most important input to the flavour tagging neural 
networks used for jets with at least two reconstructed vertices. 
If only the primary vertex is found, e.g. because the heavy flavour decay occurs so close 
to it that the two vertices cannot be resolved from each other, different variables are used 
to distinguish between jet flavours. 
In that case, the best variable identified so far is the joint probability of all tracks 
to originate from the primary vertex, which is calculated independently from the impact 
parameter significances in the $R$-$\phi$ plane and in $z$. 
It is complemented by impact parameter and impact parameter significance as well as 
momentum of the two most significant tracks in the jet. 

Altogether, the LCFIVertex code provides 9 neural networks, three each for the three cases 
that exactly one, exactly two and at least three vertices are found by \verb|ZVTOP|. 
Of the three networks for each of the cases, one is used to identify $b$-jets, one to 
identify $c$-jets in a sample with arbitrary background composition including all jet 
flavours and one to identify $c$-jets for samples for which the background is known to 
consist of $b$-jets only (this is the case for some physics processes, and permits better 
$c$-jet identification compared to the inclusive tag). 
A more detailed description of the different inputs and the relative weight of their 
contribution to the flavour tag is given elsewhere
\cite{Hawkings:2000,LCFIVertexPaperInPrep}. 
While this method forms the default approach, ensuring a high degree of flexibility has 
been a key aspect in developing the new C++ code; so network architecture, node type and 
training algorithm as well as input variables can be modified and improved or adjusted 
by the user to meet the special needs of a given analysis.

The quark charge algorithm builds strongly on the successful SLD method \cite{Thom:2002}, 
with modifications developed subsequently by LCFI \cite{Hillert:2005}. 
With the C++ implementation of the \verb|ZVKIN| algorithm \cite{Jeffery}, the foundation 
for use of the SLD charge dipole method in the ILC context has been laid; however, further 
studies will be required e.g.~to optimise \verb|ZVKIN| parameters.

Since the first release that made available these core algorithms, the package has been 
considerably extended and improved in performance. 
A module for identification and removal of tracks from $K_S$ and $\Lambda$ decays and from 
photon conversions was added (K.~Harder), and code implemented to perform a fit of the 
impact parameter significance distribution (E.~Devetak), required for calculation of one 
of the flavour tag inputs. 
Both are described in more detail elsewhere \cite{LCFIVertexPaperInPrep}. 

Technical improvements include the move of vertex charge calculation to a dedicated processor 
to decouple its implementation from the calculation of the flavour tag inputs (E.~Devetak), 
an interface to a Kalman filter by Gorbunov and Kisel (T.~Lastovicka), leading to increased 
speed of the IP fit processor, and code to ensure the package is compatible with the DST 
format recently agreed by the ILD concept group (C.~Lynch). 

Focus of current work is the optimisation of track selection and other code parameters as 
well as the training of new neural networks based on GEANT4-MC and full reconstruction 
(R.~Walsh), presented at recent meetings and conferences, see e.g. \cite{Walsh:2008}.
This work is aimed at providing a tuned code configuration for ILD detector optimisation 
towards the LoI. 

Also in view of the benchmark studies for the LoI, extensive diagnostic tools have been 
added (V. Martin), including plots of all flavour tag inputs and neural network output 
variables, each both inclusively and separately for the 1-, 2- and 3-vertex category, 
tables of efficiency, purity and corresponding neural network cut value and graphs of purity 
vs.~efficiency as well as of mistag rates for all tags. 

In terms of new algorithms, a "vertex cheater" using MC information on which tracks originate 
at the same space point and the new \verb|ZVMST| vertex finding approach have recently been 
developed (S.~Hillert).
%
%
\section{The new vertex finder algorithm ZVMST}
\label{ZVMST}
Minimum spanning trees (MSTs) are a mathematical optimisation tool with a wide range of 
applications such as in source detection in gamma ray images \cite{MSTForGammaRayDetection}, 
where this approach dates back to the early 1980s. 
As the method exploits topological information - in the astrophysical example the 
connectedness of the detected photons - it provides a natural approach to topological 
vertex finding. 
Mathematically,  minimum spanning trees are a special type of graph. 
A graph is a set of nodes connected by edges that can have weights assigned to them. 
Trees are graphs that do not contain any loops. 
For graphs with weighted edges, the minimum spanning tree is defined as the tree that 
minimizes the overall weight in the subset of graph edges selected. 
It has been mathematically proven that if no two weights are equal there always exists 
a unique solution for the minimum spanning tree.  
Efficient algorithms for finding this solution exist and optimized implementations are 
available in the graph library of the C++ package boost \cite{boost:2001}, already used 
by the LCFIVertex package. 

A central idea of the standard \verb|ZVTOP_ZVRES| algorithm is to describe each track by a 
probability density function $f_{i}(\vec{r})$ in 3D space and to use these to define a 
vertex function $V(\vec{r})$ that yields higher values in the vicinity of true vertex 
locations and lower values elsewhere, as well as providing a criterion for when two 
vertex candidates are resolved from each other.  
The definition of these functions is given and motivated in more detail elsewhere 
\cite{Jackson:1996sy}. 

In the \verb|ZVMST| algorithm, the efficient minimization of weights achieved by the 
minimum spanning tree is used to maximize the overall vertex function of selected vertex 
candidates. 
The \verb|ZVMST| algorithm has two main stages: first a small number - typically between 1 
and 5 - of 3D positions at which vertices are likely to be found is chosen on the basis of 
the vertex function. 
In the second phase tracks are assigned to these candidate vertex positions, using both 
the value of the Gaussian probability tube of each track at each of the selected space 
points and the vertex function value at these points.

To select the candidate vertex positions, the initial step is identical with that of 
\verb|ZVRES|: for all possible two-track combinations in the input jet a vertex-fit is 
attempted, and combinations discarded that have a fit-$\chi^{2}$ above a user-settable 
cut value (default: 10) or for which the vertex function at the resulting fit position 
is below 0.0001. 
The retained two-track combinations are used to set up a mathematical graph structure, 
in which each node corresponds to one of the tracks in the jet, and each edge corresponds 
to a successful vertex fit of the two tracks that it connects. 
Note that a connection is only made if the corresponding fit passes the cuts described 
above. 
As weight for the edge, the inverse of the vertex function at the vertex position 
obtained from the two-track fit is chosen.

The graph is passed as input to the minimum spanning tree algorithm. 
This algorithm selects a set of at most $N - 1$ edges for $N$ input nodes (or less if 
the input graph contains unconnected nodes) in such a way that the overall weight is 
minimised. 
In this case, because of the choice of the weights, this minimisation corresponds to 
maximising the sum of the vertex function values for the selected two-track candidate 
vertices.

Often some of the $N - 1$ selected candidates will correspond to the same physical 
vertex, especially for multi-prong vertices and the primary vertex. 
Therefore, sets of two-track candidates that correspond to one physical vertex need to 
be identified and only one optimized position derived for each set. The details of this 
stage of assigning tracks to vertices are described elsewhere \cite{Hillert:2008}. 
%
%
\section{Performance comparison of ZVMST and ZVRES at 
$\sqrt{s} = 91.2\,\mathrm{GeV}$}
\label{PerformanceComparison}
%
\begin{figure*}[h!]
\begin{center}
\begin{tabular}{cc}
\mbox{\includegraphics[width=0.45\columnwidth]{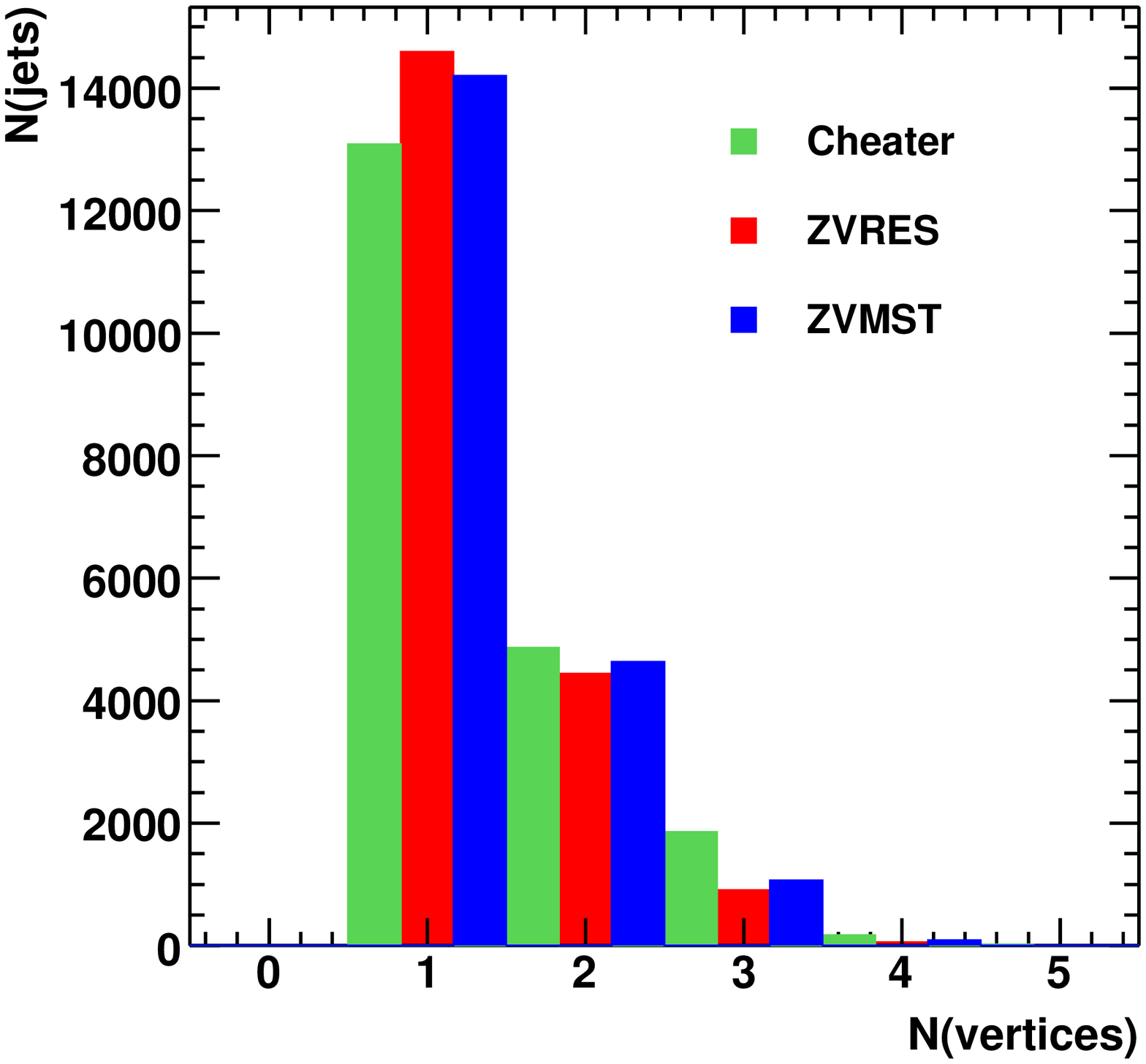}} &
\mbox{\includegraphics[width=0.45\columnwidth]{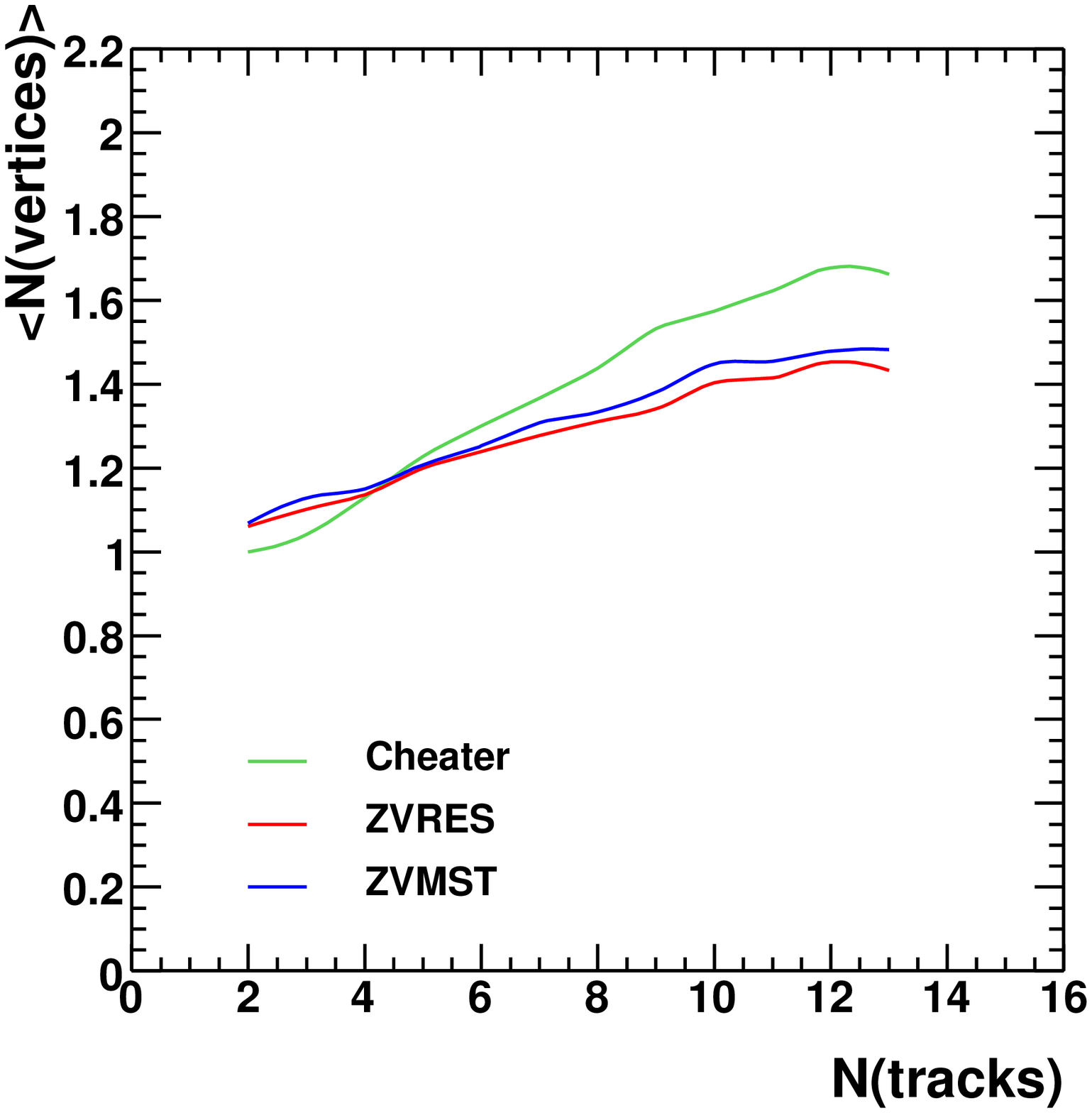}}\\
(a) & (b)\\
\end{tabular}
\end{center}
\caption{\textit{Multiplicity of vertices found by the two topological vertex reconstruction
algorithms ZVRES and ZVMST. The multiplicity of reconstructable vertices as found by
the vertex cheater is shown for comparison. (Reconstructable vertices are all vertices
that contain at least two tracks that are assigned to the same jet by the jet-finder).
Shown are (a) the inclusive distribution and (b) the average vertex multiplicity as
function of track multiplicity in the input jet.}}
\label{Fig:VertexMultiplicity}
\end{figure*}
The performance of the new vertex finder ZVMST has been studied at a centre of mass energy 
of $91.2\,\mathrm{GeV}$ and compared to that of the standard \verb|ZVTOP_ZVRES| algorithm. 
A cheater algorithm has been implemented which uses MC information to look up which tracks 
originate from the same space points, and passes these combinations through the same 
vertex fitter used for \verb|ZVRES| and \verb|ZVMST|. 
This cheater hence indicates the performance that could be achieved with perfect assignment 
of tracks to vertices. 

Fig.~\ref{Fig:VertexMultiplicity} shows the inclusive vertex multiplicity and the vertex 
multiplicity as function of the number of tracks corresponding to the vertex, for 
\verb|ZVMST|, \verb|ZVRES| and the cheater. 
Both vertex finding algorithms yield a smaller number of vertices than the cheater, with 
\verb|ZVMST| finding a slightly larger number than \verb|ZVRES|. 
In particular as the track multiplicity increases, the number of vertices falls short of 
the true MC vertex multiplicity by an increasing amount. 
At high multiplicity, \verb|ZVMST| is closer to the MC truth, while this is the case for 
\verb|ZVRES| 
at low multiplicity; the cross-over point is at a track multiplicity of about 4. 

\begin{figure*}[h]
\centering
\mbox{
\includegraphics[width=0.95\columnwidth]{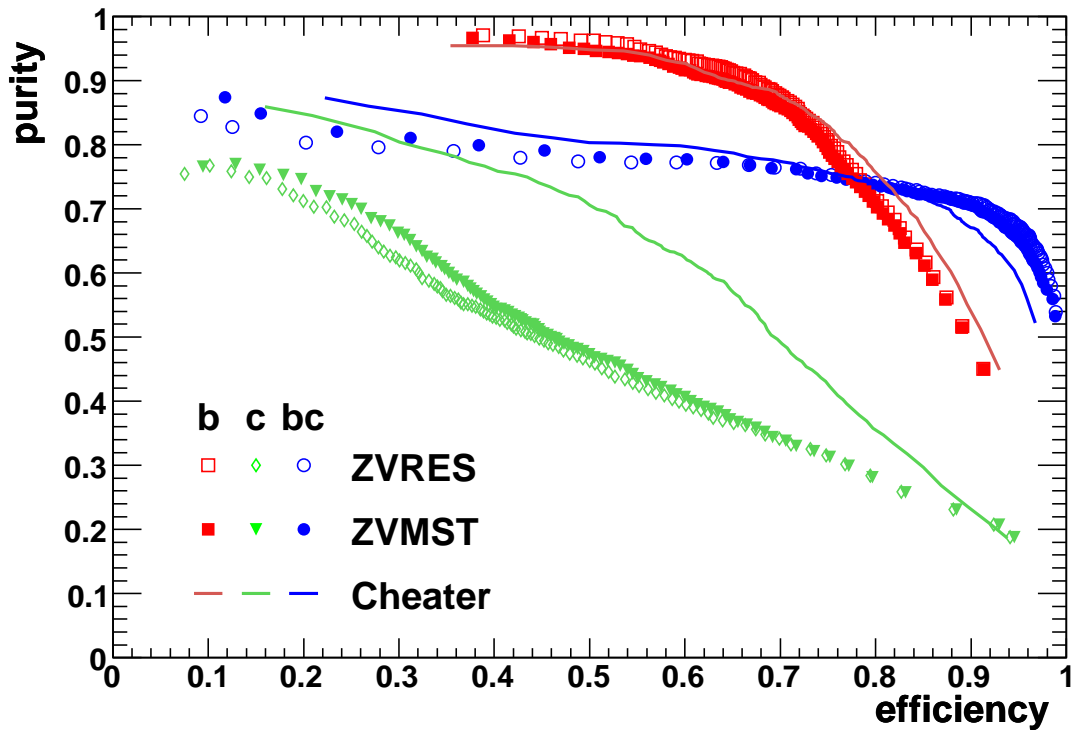}}
\caption{\textit{Comparison of tagging performance at the $Z$-resonance obtained using the
new ZVMST vertex finder compared to results obtained using ZVTOP's ZVRES algorithm
and for a vertex cheater using MC information for track-to-vertex assignment.
Tagging purity is shown as function of efficiency for $b$-jets
and $c$-jets. Performance for $c$-jets assuming only $b$-background (labelled ``bc'') is also
shown. }}
\label{Fig:FlavourTagPerformance}
\end{figure*}

The performance of the track to vertex assignment for the two algorithms was studied and 
is discussed in detail elsewhere \cite{Hillert:2008}. 
It yields similar performance for both algorithms, with some aspects being better for one, 
and others being better for the other.

Fig.~\ref{Fig:FlavourTagPerformance} shows the flavour tag performance in terms of purity 
vs efficiency for a $91.2\,\mathrm{GeV}$ mixed sample with natural fractions of $b$-, $c$- 
and light jets. 
The result obtained from the \verb|ZVMST| algorithm is compared to the performance of 
\verb|ZVRES| and the cheater. 
The new \verb|ZVMST| algorithm yields an improved $c$-tag purity over the full efficiency 
range, with the difference compared to \verb|ZVRES| reaching values up to $5\%$ at low 
efficiency. 
However, $b$-tag purity is degraded by up to $1.5\%$ compared to \verb|ZVRES|. 
This is consistent with the trend found during the code development that the $c$-tag is 
the most sensitive of the three tags whenever a change is made e.g.~to track selection, 
track reconstruction, detector geometry etc.  
The comparison with the cheater performance shows that if improvements could be made to 
the track-to-vertex assignment this should directly result in an improvement in $c$-tag 
performance.

It should be noted that  all three results are found using the flavour tag neural 
networks that were trained with the fast MC SGV at an earlier stage of code development. 
Since the time when the \verb|ZVMST| study has been performed it was shown that flavour tag 
performance is sensitive to the networks used \cite{Walsh:2008}, and that a more realistic 
comparison would therefore require training dedicated networks to be used with each of 
the algorithms. 
This could also explain why in Fig. 2 the $b$-tag purity obtained from the cheater is 
slightly worse than that obtained from the two realistic vertex finders.
%
%
\section{Summary and conclusions}
\label{SummaryAndConclusions}
The LCFIVertex package, providing vertexing, flavour tagging and quark charge 
reconstruction, is an essential tool for the preparation of detector LoIs for the ILC, 
with generic applicability to high-precision vertex detectors. 
Additions to the core functionality of the code provided by the first release in 2007 
include extensive diagnostics, technical improvements and work towards a default 
configuration of the code, as well as the new vertex finder \verb|ZVMST|, based on a 
minimum spanning tree approach.

In a performance study of the \verb|ZVMST| algorithm at $\sqrt{s} = 91.2\,\mathrm{GeV}$, 
the algorithm has been shown to be competitive with the leading algorithm at the ILC, 
\verb|ZVTOP_ZVRES|. 
Vertex multiplicities from \verb|ZVMST| are slightly closer to the reference values 
obtained from MC truth track combinations than is the case for \verb|ZVRES|. 
Track-to-vertex assignment is similar for both algorithms, some aspects being improved 
for \verb|ZVMST|, while others are degraded. 

At $\sqrt{s} = 91.2\,\mathrm{GeV}$ flavour tag performance has been compared using the 
flavour tag networks obtained from the fast MC SGV, which have subsequently been shown 
not to be optimal for any of the algorithms if using GEANT4 based MC and realistic event 
reconstruction. 
Within these boundary conditions, \verb|ZVMST| yields an up to $5\%$ increase in $c$-tag 
purity and an up to $1.5\%$ degradation in $b$-tag purity compared to \verb|ZVTOP_ZVRES|. 
A further limitation of this study is the fact that the code parameters used are not 
yet optimized; though \verb|ZVRES| parameters are set to the values obtained from an 
earlier GEANT3-based study and a preliminary parameter optimization at $91.2\,\mathrm{GeV}$ 
was performed for \verb|ZVMST| \cite{Hillert:2008}.

%

\end{document}